\documentclass[a4paper,10pt,twoside]{cpc-hepnp}
\usepackage{CJK,upgreek,fancyhdr}
\usepackage{multicol}
\usepackage{graphicx}
\usepackage{booktabs}
\usepackage{amssymb,bm,mathrsfs,bbm,amscd}
\usepackage[tbtags]{amsmath}
\usepackage{lastpage}

\begin{document}
\begin{CJK*}{GB}{gbsn}

\fancyhead[c]{\small Chinese Physics C~~~Vol. xx, No. x (201x) xxxxxx}
\fancyfoot[C]{\small 010201-\thepage}

\footnotetext[0]{Received 31 June 2015}

\title{Sub-leading flow modes in PbPb collisions \\ 
at $\sqrt{s_{NN}}$ = 2.76~TeV from HYDJET++ model\thanks{Supported by Ministry of Education, Science and Technological Development of the Republic of Serbia (171019) }}

\author{
      P. Cirkovic$^{1}$
\quad D. Devetak$^{2}$
\quad M. Dordevic$^{2}$
\quad J. Milosevic$^{2,3}$\email{Jovan.Milosevic@cern.ch}
\quad M. Stojanovic$^{2}$
}
\maketitle

\address{
$^1$ University of Belgrade and Institute of physics, P.O. Box 68, 11081 Belgrade, Serbia\\
$^2$ University of Belgrade and Vin\v{c}a Institute of Nuclear Sciences, P.O. Box 522, 11001 Belgrade, Serbia\\
$^3$ \emph{Also at: University of Oslo, Department of Physics, Oslo, Norway}
}

\begin{abstract}
Recent LHC results on the appearance of sub-leading flow modes in PbPb collisions at 2.76~TeV, related to initial-state fluctuations, are analyzed and interpreted within the HYDJET++ model. Using the newly introduced Principal Component Analysis (PCA) method applied to two-particle azimuthal correlations extracted from the model calculations, the leading and the sub-leading flow modes are studied as a function of the transverse momentum ($p_{T}$) over a wide centrality range. The leading modes of the elliptic ($v^{(1)}_{2}$) and triangular ($v^{(1)}_{3}$) flow calculated within the HYDJET++ model reproduce rather well the $v_{2}\{2\}$ and $v_{3}\{2\}$ coefficients experimentally measured using the two-particle correlations. Within the $p_{T} \le $~3~GeV/c range where hydrodynamics dominates, the sub-leading flow effects are greatest at the highest $p_{T}$ of around 3~GeV/c. The sub-leading elliptic flow mode ($v^{(2)}_{2}$), which corresponds to $n = 2$ harmonic, has a small non-zero value and slowly increases from central to peripheral collisions, while the sub-leading triangular flow mode ($v^{(2)}_{3}$), which corresponds to $n = 3$ harmonic, is even smaller and does not depend on centrality. For $n = $~2, the relative magnitude of the effect measured with respect to the leading flow mode shows a shallow minimum for semi-central collisions and increases for very central and for peripheral collisions. For $n = $~3 case, there is no centrality dependence. The sub-leading flow mode results obtained from the HYDJET++ model are in a rather good agreement with the experimental measurements of the CMS Collaboration.
\end{abstract}

\begin{keyword}
Hydrodynamics flow, Initial-state fluctuations, Principal Component Analysis, HYDJET++
\end{keyword}

\begin{pacs}
25.75.Gz, 25.75.Dw
\end{pacs}

\footnotetext[0]{\hspace*{-3mm}\raisebox{0.3ex}{$\scriptstyle\copyright$}2013
Chinese Physical Society and the Institute of High Energy Physics
of the Chinese Academy of Sciences and the Institute
of Modern Physics of the Chinese Academy of Sciences and IOP Publishing Ltd}%

\begin{multicols}{2}

\section{Introduction}

\label{intro}
According to Quantum Chromodynamics, at sufficiently high energy density which can be achieved in ultra-relativistic heavy-ion collisions, a new state of matter, called Quark-Gluon-Plasma (QGP), is created. One of main features of the QGP is its collective expansion which could be described by relativistic hydrodynamic flows. Due to the different pressure gradients in different directions, the initial spatial eccentricity converts into momentum anisotropy, observed in the final state as a preferential emission of particles in a certain plane.

Quantitatively, the anisotropic hydrodynamic flow is described by Fourier decomposition of the hadron yield distribution in azimuthal angle, $\phi$,~\cite{Ollitrault:1993ba,Voloshin:1994mz,Poskanzer:1998yz}
\begin{equation}
\label{F1}
\frac{dN}{d\phi} \propto 1+2\sum_{n}v_{n}\cos[n(\phi - \Psi_{n})],
\end{equation}
where Fourier coefficients, $v_{n}$, characterize magnitude of the azimuthal anisotropy measured with respect to the flow symmetry plane angle, $\Psi_{n}$. The angle $\Psi_{n}$ determines the direction of maximum final-state particle density and can be reconstructed from the emitted particles themselves. The most analyzed anisotropic flow is the second order Fourier coefficient, $v_{2}$, called elliptic flow. The angle $\Psi_{2}$ corresponds to the flow symmetry plane which is correlated with the participant plane spanned over the beam direction and the shorter axis of the approximately elliptical shape of the nucleon overlap region. Due to the initial-state fluctuations of the position of nucleons at the moment of impact, higher-order deformations of the initial geometry are induced, which lead to the appearance of higher-order Fourier harmonics ($v_{n}$, $n \ge $3 in Eq.~(\ref{F1})). They are measured with respect to the corresponding flow symmetry plane angles, $\Psi_{n}$~\cite{Alver:2010gr}. The collective behavior of a strongly-coupled hot and dense QGP has been studied using the azimuthal anisotropy of emitted particles detected at experiments at the Relativistic Heavy Ion Collider (RHIC)~\cite{Back:2002gz,Ackermann:2000tr,Adcox:2002ms}. The studies have been continued also with the experiments~\cite{Aamodt:2010pa,ALICE:2011ab,Abelev:2014pua,Adam:2016izf,ATLAS:2011ah,ATLAS:2012at,Aad:2013xma,Chatrchyan:2012wg,Chatrchyan:2012ta,Chatrchyan:2013kba,CMS:2013bza,Khachatryan:2015oea} at the Large Hadron Collider (LHC) where significantly higher collision energies are achieved.

Another experimental method to determined the $v_{n}$ coefficients uses two-particle azimuthal correlations~\cite{Wang:1991qh}. These correlations can be also Fourier decomposed as
\begin{equation}
\label{F2}
\frac{dN^{pair}}{d\Delta\phi} \propto 1+2\sum_{n}V_{n\Delta}\cos(n\Delta\phi),
\end{equation}
where $\Delta\phi$ is a relative azimuthal angle of a particle pair. The two-particle Fourier coefficient $V_{n\Delta}$ is expected to factorize as
\begin{equation}
\label{Fact}
V_{n\Delta}(p^{a}_{T},p^{b}_{T}) = v_{n}(p^{a}_{T})v_{n}(p^{b}_{T}),
\end{equation}
into a product of the anisotropy harmonics $v_{n}$. 

A key assumption for correctness of Eq.~(\ref{Fact}) is that the flow symmetry plane angle $\Psi_{n}$ in Eq.~(\ref{F1}) is a global quantity for a given collision. The effect has been theoretically predicted in~\cite{Gardim:2012im,Heinz:2013bua}. It is shown that even if the hydrodynamic flow is the only source of the two-particle correlations, initial-state fluctuations turns the flow symmetry plane from a global to both, $p_{T}$ and $\eta$ dependent quantity. Lumpy hot-spots raised from the initial-state fluctuations can generate a local pressure gradient which makes the corresponding local flow symmetry plane to be slightly different but still correlated with the global flow symmetry plane. This effect of initial-state fluctuations thus breaks the factorization relation of Eq.~(\ref{Fact}). A significant breakdown of the factorization assumption expressed through Eq.~(\ref{Fact}) has been observed both in the transversal $p_{T}$ and longitudinal $\eta$ direction\footnote{Pseudorapidity $\eta$ is defined as $-ln\tan(\theta/2)$ where $\theta$ is the polar angle.} in symmetric PbPb collisions~\cite{CMS:2013bza,Khachatryan:2015oea,Zhou:2014bba} as well as in asymmetric pPb collisions~\cite{Khachatryan:2015oea,Zhou:2014bba}.

Recently, a new approach which employs the Principal Component Analysis (PCA) to study the flow phenomena is introduced~\cite{Bhalerao:2014mua,Mazeliauskas:2015vea}. Using a PCA approach, $V_{n\Delta}$ coefficients of the observed two-particle azimuthal correlations as a function of both particles $p_{T}$ are represented through the leading and the sub-leading flow mode terms. The leading flow modes are essentially equivalent to anisotropy harmonics ($v_{n}\{2\}$) extracted from two-particle correlation methods. As a consequence of initial-state fluctuations, the sub-leading flow modes could appear as the largest sources of factorization breaking. The PCA study of this effect can give new insights into the expansion dynamics of the strongly coupled QGP, and serves as an excellent tool for testing the hydrodynamical models. 

This paper is organized in a following way. The basic features of HYDJET++ model~\cite{Lokhtin:2008xi} are described in Sect.~2. Details of the applied construction of the two-particle correlation functions, as well as the PCA approach in flow analysis are given in Sect.~3. Using HYDJET++ model, approximately 40M PbPb collisions at $\sqrt{s_{NN}} = $2.76~TeV are simulated and analyzed. The obtained results together with the corresponding discussions are given in Sect.~4. The results are presented over a wide range of centralities going from ultra central (0-0.2\% centrality\footnote{The centrality in heavy ion collisions is defined as a fraction of the total inelastic PbPb cross section, with 0\% denoting the most central collisions.}) up to peripheral (50-60\% centrality) PbPb collisions. The analyzed $p_{T}$ interval is restricted to $p_{T} \le $~3~GeV/c range where hydrodynamics dominates. A disscussion concerning the results obtained under different HYDJET++ model switches is given in Sect.~5. Conclusions are given in Sect.~6.

\section{HYDJET++}
\label{HYDJET}
The Monte Carlo HYDJET++ model simulates relativistic heavy ion collisions in an event-by-event manner. It is made of two components which simulate soft and hard processes. The soft part provides the hydrodynamical evolution of the system while the hard part describes multiparton fragmentation within the formed medium. Within the hard part, jet quenching effects are also taken into account. The minimal transverse momentum $p^{min}_{T}$ of hard scattering of the incoming partons regulates does it would contribute to the soft or to the hard part. The partons which are produced with $p_{T} < p^{min}_{T}$, or which are quenched below $p^{min}_{T}$ do not contribute to the hard part. The hard part of the model consists of PYTHIA~\cite{Sjostrand:2006za} and PYQUEN~\cite{Lokhtin:2005px} event generators. These generators simulate initial parton-parton collisions, radiative energy loss of partons and parton hadronization. Within the soft part of the HYDJET++ model, the magnitude of the elliptic flow is regulated by spatial anisotropy $\epsilon(b)$ which is the elliptic modulation of the final freeze-out hyper-surface at a given impact parameter vector\footnote{In an ideal circle-like geometry, impact parameter $\vec b$ is a vector which connects centers of the colliding nuclei.} magnitude $b$, and by momentum anisotropy $\delta(b)$ which gives the modulation of the flow velocity profile. Additionally introduced triangular modulation of the freeze-out hyper-surface, $\epsilon_{3}$, determines the $v_{3}$ magnitude. The events can be generated under several switches. The most realistic one, 'flow+quenched jets', includes both hydrodynamics expansion and quenched jets. In this analysis, the pure 'flow' switch is also used. The details of the model can be found in the HYDJET++ manual~\cite{Lokhtin:2008xi}.

\section{Prescription of the Principle Component Analysis technique}
\label{sec:ana}
\subsection{Two-particle correlation function}
\label{sec:2-part}
The construction of the two-dimensional two-particle correlation function follows the definition adopted within the CMS experiment. Any charged pion from the $|\eta| < $ 2 range can be used as a 'trigger' particle. In order to perform a differential analysis, all events are divided into eight centrality classes, while the analyzed $p_{T}$ range has seven non-equidistant intervals. Since in an event there can be more than one trigger particle from a given $p_{T}$ interval, the corresponding total number is denoted by $N_{trig}$. In each event, every trigger particle is paired with all of the remaining charged pions from the $|\eta| < $ 2 range within a given $p_{T}$ interval. The signal distribution, $S(\Delta\eta, \Delta\phi)$, is defined as the yield of the per-trigger-particle pairs within the same event,
\begin{equation}
\label{S}
S(\Delta\eta,\Delta\phi) = \frac{1}{N_{trig}}\frac{d^{2}N^{same}}{d\Delta\eta d\Delta\phi}.
\end{equation}
In Eq.~(\ref{S}), $N^{same}$ denotes the per-trigger-particle pairs yield within a given ($\Delta\eta,\Delta\phi$) bin where $\Delta\eta$ and $\Delta\phi$ are corresponding differences in pseudorapidity and azimuthal angle between the two charged pions which are forming the pair. The background distribution, denoted with $B(\Delta\eta, \Delta\phi)$, is constructed using the technique of mixing topologically similar events which ensure that the pairs are not physically correlated. Here, topological similarity means that events which are mixed have relative difference in multiplicity smaller than 5\%. The trigger particles from one event are combined (mixed) with all of the associated particles from a different event. In order to reduce contribution to the statistical uncertainty from the background distribution, associated particles from 10 randomly chosen events are used. In the background distribution, defined as
\begin{equation}
\label{B}
B(\Delta\eta,\Delta\phi) = \frac{1}{N_{trig}}\frac{d^{2}N^{mix}}{d\Delta\eta d\Delta\phi},
\end{equation}
$N^{mix}$ denotes the number of mixed-event pairs in a given ($\Delta\eta,\Delta\phi$) bin. Due to the fact that pairs are formed from uncorrelated particles, the background gives a distribution of independent particle emission.

The two-dimensional two-particle differential correlation function is then defined as the normalized ratio of the {\it signal} to the {\it background} distribution
\begin{equation}
\label{2Dcorr}
\frac{1}{N_{trig}}\frac{d^{2}N^{pair}}{d\Delta\eta d\Delta\phi} = B(0,0)\frac{S(\Delta\eta,\Delta\phi)}{B(\Delta\eta,\Delta\phi)}.
\end{equation}
The normalization factor, B(0, 0), is the value of the background distribution at $\Delta\eta = $ 0 and $\Delta\phi = $ 0.

In order to obtain azimuthal anisotropy harmonics, $v_{n}\{2\}$, the projection of the two-dimensional correlation function given by Eq.~(\ref{2Dcorr}) onto $\Delta\phi$ axis can be Fourier decomposed as given in Eq.~(\ref{F2}). In order to suppress the short-range correlations arising from jet fragmentation and resonance decays, an averaging over $|\Delta\eta| > $ 2  is applied. This is one way to extract two-particle Fourier coefficients $V_{n\Delta}$ introduced in Eq.~(\ref{F2}).
\subsection{Principle Component Analysis}
\label{sec:PCA}
PCA is a statistical method that orders fluctuations in data by size or so-called components. Application of this method in frames of anisotropic flow was introduced in~\cite{Bhalerao:2014mua} and further investigated in \cite{Mazeliauskas:2015vea,Mazeliauskas:2015efa}. By extracting principal components from the two-particle correlation data one can probe the presence of any event-by-event flow fluctuations.

Section 3.1 shows how the two-particle Fourier harmonics $V_{n\Delta}$ are extracted using the fitting procedure. An alternative approach for calculating the Fourier harmonics $V_{n\Delta}$ is applied in~\cite{CMS:2013bza} as,
\begin{equation}
\label{2PFC}
V_{n\Delta}(p^{a}_{T},p^{b}_{T}) = \langle\langle \cos(n\Delta\phi) \rangle\rangle_{S} - \langle\langle \cos(n\Delta\phi) \rangle\rangle_{B},
\end{equation} 
where $S$ and $B$ stands for the {\it signal} and for the {\it background}, respectively. Here, double brackets $\langle\langle \cdot\rangle\rangle$ denote averaging over charged pion pairs and over all events from the given centrality class. The procedure of forming pairs in $S$ and $B$, with the pseudo-rapidity cut $|\Delta\eta|>2$, is identical as in the fitting case. Following the procedure given in~\cite{Bhalerao:2014mua}, in order to use the PCA technique a single bracket definition for the two-particle Fourier harmonics $V^{PCA}_{n\Delta}$ is used,
\begin{equation}
\label{2PCA}
V^{PCA}_{n\Delta}(p^{a}_{T},p^{b}_{T}) = \langle \cos(n\Delta\phi) \rangle_{S} - \langle \cos(n\Delta\phi) \rangle_{B},
\end{equation}
where $\langle\cdot\rangle$ refers to averaging over all events from the given centrality class. The PCA method is applied by doing the eigenvalue decomposition of the covariance matrix that is built out of the $V^{PCA}_{n\Delta}$ harmonics.  By defining $N_{b}$ differential $p_{T}$ bins one can construct the corresponding covariance matrix $[\hat{V}(p^a_T,p^b_T)]_{N_{b}{\times}N_{b}}$. The diagonal elements are harmonics with correlated particles $a$ and $b$ taken from the same $p_{T}$ bin and the non-diagonal elements are harmonics  with correlated particles $a$ and $b$ taken from the different $p_{T}$ bins. In this analysis the $p_{T}$ range, going from 0.3 to 3.0~GeV/c, has been divided into $N_{b} = $~7 non-equidistant $p_{T}$ bins. By solving the eigenvalue problem of the $[\hat{V}(p^a_T,p^b_T)]_{N_{b}{\times}N_{b}}$ matrix, a set of the eigenvalues, $\lambda^{(\alpha)}$, and eigenvectors, $e^{(\alpha)}$, has been obtained. Here, $\alpha = 1,...,N_{b}$. A new $p_{T}$-dependent observable, $V^{(\alpha)}_{n}(p_{T})$, is introduced as
\begin{equation}
\label{Fmode}
V^{(\alpha)}_{n}(p_{T}) = \sqrt{\lambda^{(\alpha)}}e^{(\alpha)}(p_{T}),
\end{equation}
referring to it as mode for the given $\alpha$. The first mode (denoted with $\alpha = $1) corresponds to the first greatest variance of data, the second mode (denoted with $\alpha = $2) corresponds to the second greatest variance of data and so on. The modes are not of the same order as the standard $v_n\{2\}$ harmonics and a normalized observable is defined as,
\begin{equation}
\label{spFmode}
v^{(\alpha)}_{n}(p_{T}) = \frac{V^{(\alpha)}_{n}(p_{T})}{\langle M(p_{T}) \rangle},
\end{equation}
where $\langle M(p_{T}) \rangle$ denotes the averaged multiplicity in a given $p_{T}$ bin. The multiplicity normalization, introduced in~\cite{Bhalerao:2014mua}, follows from the fact that the two-particle harmonics from Eq.~(\ref{2PFC}) and Eq.~(\ref{2PCA}) differ by a factor of ${\langle{N^{pairs}(p_T,p_T)}\rangle}\simeq{{\langle}M(p_T)\rangle}^2$. However, the last equality is broken when the pseudo-rapidity cut $|\Delta\eta|>2$ is applied. Thus, the multiplicity normalization is restored by correcting the PCA harmonics from Eq.~(\ref{2PCA}) as $V^{PCA}_{n\Delta}(p^a_T,p^b_T){\mapsto}{\frac{N^{\mathrm{pairs}}(p^a_T,p^b_T,|\eta|<2.4)}{N^{\mathrm{pairs}}(p^a_T,p^b_T,|\Delta\eta|>2)}}V^{PCA}_{n\Delta}(p^a_T,p^b_T)$. The observables from Eq.~(\ref{spFmode}) for $\alpha=1$ and $\alpha=2$ will be referred to as the leading and the sub-leading flow modes respectively. The magnitude of the leading flow mode, $v^{(1)}_{n}$, should be practically equal to the $v_{n}\{2\}$ measured using the two-particle correlation method. The CMS Collaboration showed in~\cite{HIN-15-010,Milosevic:2016tiw} that the $p_{T}$ dependence of the leading elliptic and triangular flow modes for pPb collisions at 5.02~TeV and for PbPb collisions at 2.76~TeV data are in excellent agreement with corresponding two-particle measurements presented in~\cite{Chatrchyan:2013nka} and in~\cite{Aamodt:2011by}, respectively.

\section{Results}
\label{sec:res}
In order to check the consistency of extracted azimuthal anisotropies, $v_{n}$, using the method of PCA and one of the standard approaches like Fourier decomposition given by Eq.~(\ref{F2}), as well as to perform a PCA analysis in order to extract the leading and sub-leading flow modes, the two-particle correlation functions defined by the Eq.~(\ref{2Dcorr}) are constructed. For each centrality interval, ranged from the ultra-central 0-0.2\% till peripheral 50-60\%, two-particle correlation functions for 7 $p_T$ intervals between 0.3 and 3.0~GeV/c are formed. Thus, 7 diagonal and 21 non-diagonal two-particle correlation functions are produced. As examples, in Fig.~\ref{fig:1} are shown two-dimensional, in $\Delta\eta$ and $\Delta\phi$, two-particle correlation functions from HYDJET++ PbPb simulations at $\sqrt{s_{NN}} = $2.76~TeV where both particles belongs to 0.3 $ < p_{T} < $ 0.5~GeV/c (left column) and 1.5 $ < p_{T} < $ 2.0~GeV/c (middle column) interval (diagonal elements), while in the right column one particle belongs to 0.3 $ < p_{T} < $ 0.5~GeV/c and another one to 1.5 $ < p_{T} < $ 2.0~GeV/c (non-diagonal element).
\end{multicols}
\ruleup
\begin{center}
  \includegraphics[width=0.96\textwidth]{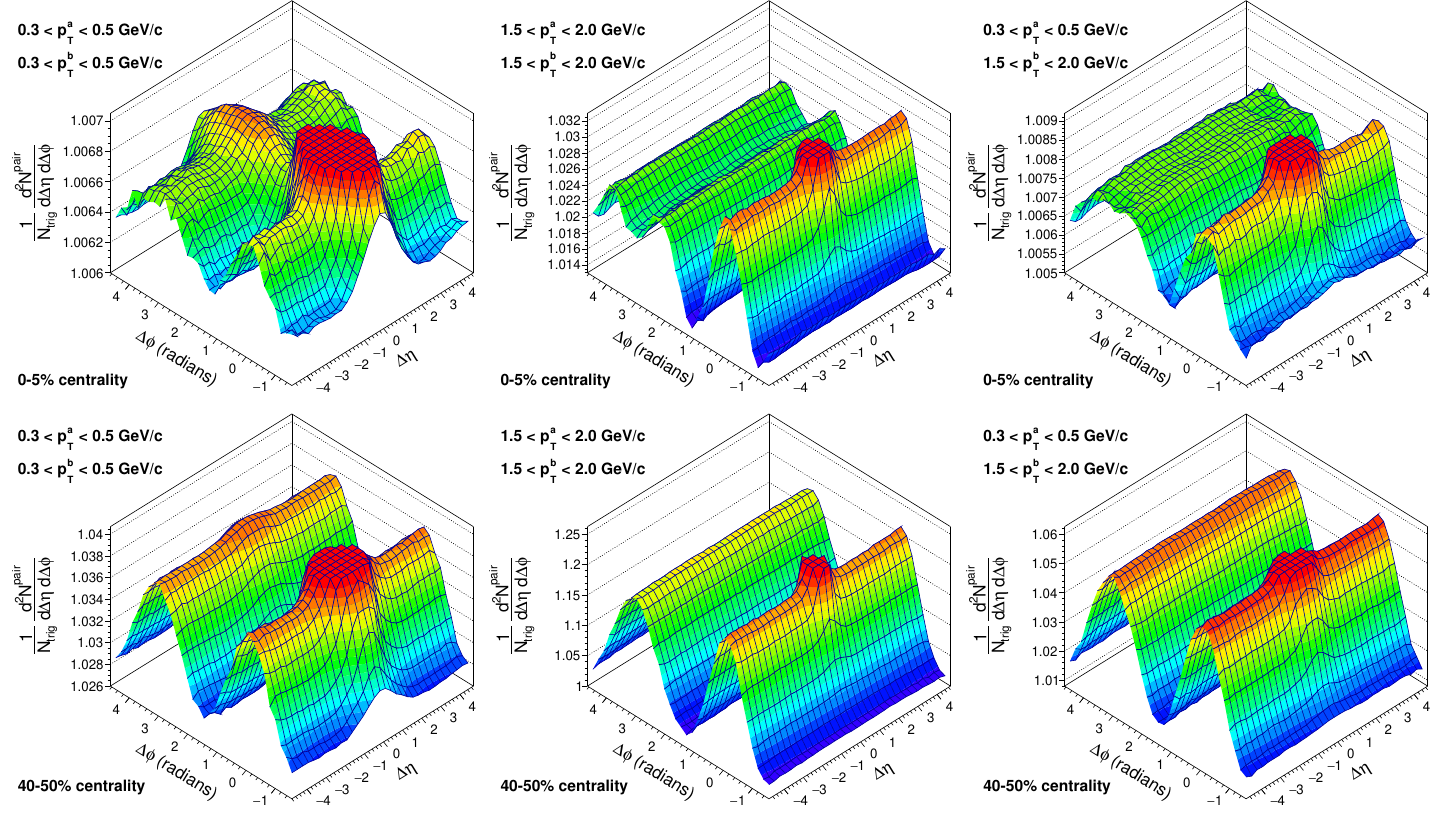}
\figcaption{\label{fig:1} Two dimensional, in $\Delta\eta$ and $\Delta\phi$, two-particle correlation functions where both particles belongs to 0.3 $ < p_{T} < $ 0.5~GeV/c (left column), 1.5 $ < p_{T} < $ 2.0~GeV/c (middle column) and the case where one particle belongs to 0.3 $ < p_{T} < $ 0.5~GeV/c and another one to 1.5 $ < p_{T} < $ 2.0~GeV/c (right column). Top (bottom) row: The correlation functions are constructed from very central 0-5\% (peripheral 40-50\%) 2.76 TeV PbPb collisions simulated within HYDJET++ model under the 'flow + quenched jets' switch.}
\end{center}
\ruledown
\begin{multicols}{2}

\noindent{The correlation functions in the top row are constructed from very central 0-5\% collisions, and those from peripheral 40-50\% collisions are presented in the bottom row. As this analysis deals with the long-range correlations, the near side peak is truncated. One can see that, beside the short-range correlated near side peak, HYDJET++ model can reproduce rather well features of the elliptic and triangular flow. For higher transverse momenta (1.5~$< p_{T} <$~2.0~GeV/c), in difference of peripheral collisions where the elliptic flow dominates, in central collisions (0-5\% centrality) the magnitude of the $v_{3}$ becomes similar to the magnitude of the $v_{2}$, and thus a clear double-bump peak is seen at the away side.}

In Fig.~\ref{fig:2} are shown the PCA results on the leading and sub-leading flow modes for the second harmonic in 8 centrality regions ranged from ultra-central (0-0.2\%) to peripheral (50-60\%) PbPb collisions at $\sqrt{s_{NN}} =$~2.76 TeV simulated within HYDJET++ event generator. The leading flow mode, $v_{2}^{(1)}$, is dominant and rather well describes the experimentally measured $v_{2}\{2\}$ from two-particle correlations taken from~\cite{CMS:2013bza} and~\cite{Aamodt:2011by}. Additionaly, due to consistency, in Fig.~\ref{fig:2} are also shown $v_{2}\{2,|\Delta\eta| > 2\}$ values measured using two-particle correlations constructed from the same HYDJET++ generated data. In Fig.~\ref{fig:2} these results are depicted with crosses and show an excelent agreement with $v_{2}^{(1)}$ extracted using the PCA method. The extracted $v_{2}^{(1)}$ has expected centrality behavior: a small magnitude at ultra-central collisions which then gradually increases going to peripheral collisions. The newly observed sub-leading flow mode of second order harmonic, $v^{(2)}_{2}$, is practicaly equal to zero at small-$p_{T}$ for all centrality bins. For $p_{T} > $~2 GeV/c, the sub-leading flow mode has a small positive value and slowly increases going from semi-central to peripheral PbPb collisions. The CMS collaboration presented in~\cite{HIN-15-010,Milosevic:2016tiw} experimentally measured the leading and sub-leading flow mode in PbPb collisions within the same $p_{T}$ range and for the same centrality bins as it is adopted in this analysis. Beside the leading flow mode, HYDJET++ predictions for the sub-leading flow mode are also in a qualitative agreement with the experimental findings from~\cite{HIN-15-010,Milosevic:2016tiw}. For centralities above 30\%, the $v^{(2)}_{2}$ magnitudes predicted by HYDJET++ model are slightly larger with respect to the ones observed from the experimental data.
\end{multicols}
\ruleup
\begin{center}
  \includegraphics[width=0.8\textwidth]{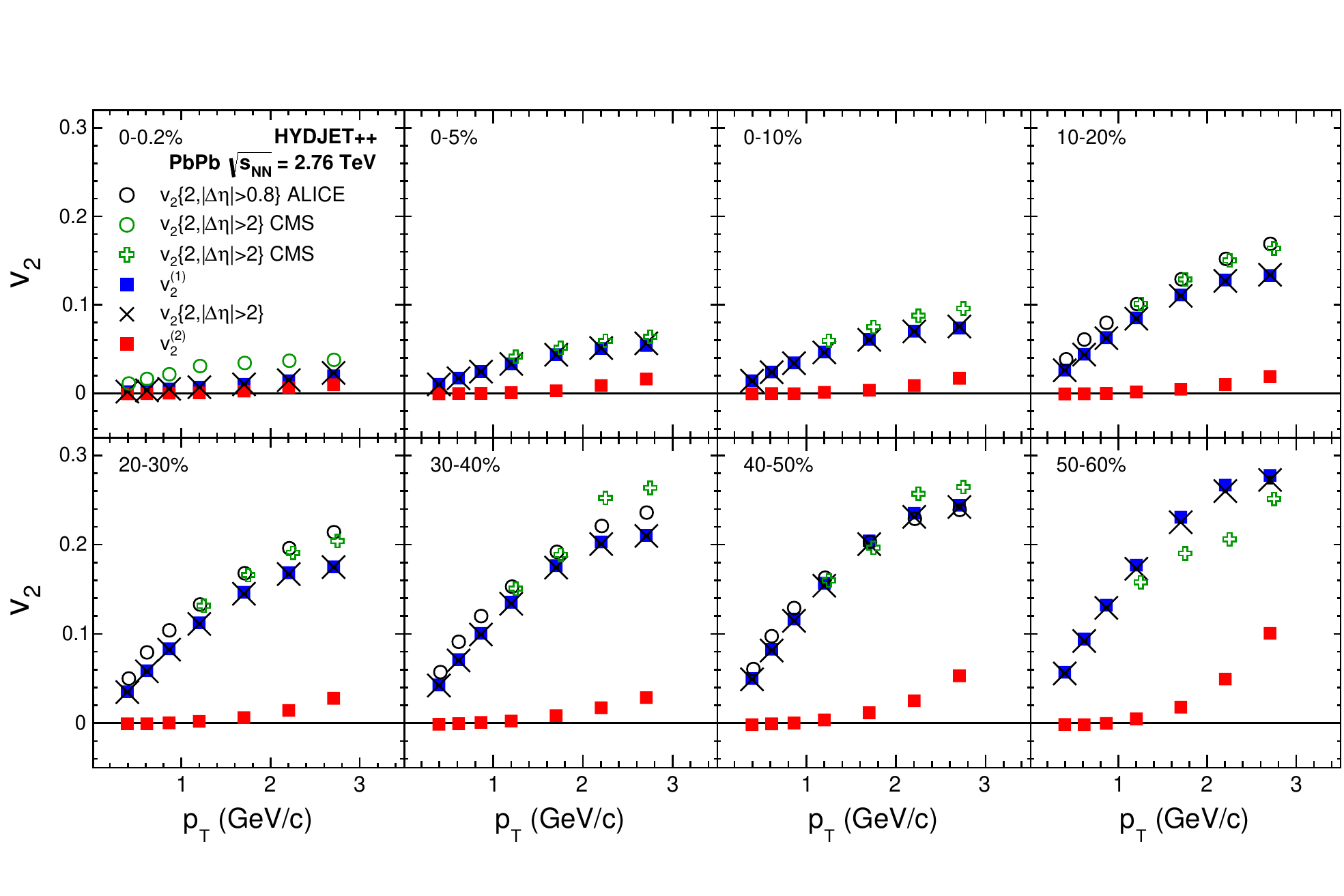}
\figcaption{\label{fig:2} The leading ($\alpha$ = 1) and the sub-leading ($\alpha$ = 2) flow mode for $n$ = 2 harmonic as a function of $p_{T}$ measured using the PCA approach in a wide centrality range of PbPb collisions at 2.76~TeV generated within the HYDJET++ model. The $v^{(1)}_{2}$ results are compared to the $v_{2}\{2\}$ measured by the CMS~\cite{CMS:2013bza} (open green circles) and~\cite{Chatrchyan:2012wg} (open green crosses) and by ALICE~\cite{Aamodt:2011by} collaborations, and to the $v_{2}\{2,|\Delta\eta| > 2\}$ extracted from the same HYDJET++ simulation using the two-particle correlation method. The error bars correspond to statistical uncertainties.}
\end{center}
\ruledown
\begin{multicols}{2}

Similarly as in Fig.~\ref{fig:2}, in Fig.~\ref{fig:3} are shown the PCA leading and sub-leading flow mode predictions of HYDJET++ model for the third harmonic. Again, the results are extracted from the 8 centrality regions, same as in Fig.~\ref{fig:2}, of PbPb collisions at $\sqrt{s_{NN}} =$~2.76~TeV. The $v^{(1)}_{3}$ is in a rather good agreement with the $v_{3}\{2\}$ results measured using two-particle correlations taken from~\cite{CMS:2013bza} and~\cite{Aamodt:2011by}, except in the case of ultra-central collisions. Also, the $v_{3}\{2\}$ extracted from the two-particle correlations formed from the same HYDJET++ generated data are in an excellent agreement with the $v^{(1)}_{3}$ obtained from the PCA method. The sub-leading mode is, up to 3~GeV/c, almost equal to zero. This supports finding from~\cite{Khachatryan:2015oea,Zhou:2014bba} that the third harmonic factorizes better than the second one. Also, the small $v^{(2)}_{3}$ values extracted from HYDJET++ simulated PbPb events are in an agreement with those found in~\cite{HIN-15-010,Milosevic:2016tiw} extracted from the experimental PbPb data.
\end{multicols}
\ruleup
\begin{center}
  \includegraphics[width=0.8\textwidth]{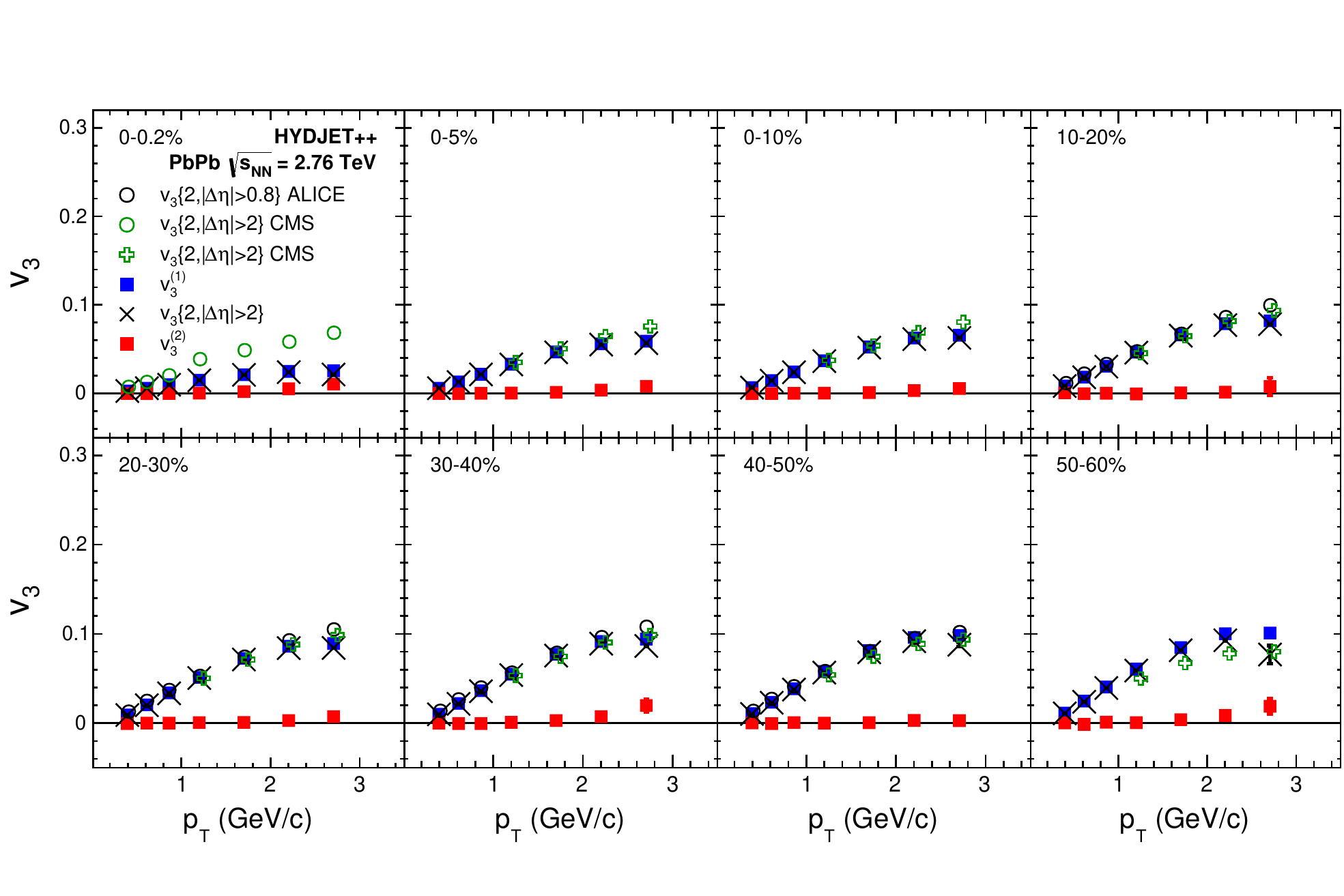}
\figcaption{\label{fig:3} The leading ($\alpha$ = 1) and the sub-leading ($\alpha$ = 2) flow mode for $n$ = 3 harmonic as a function of $p_{T}$ measured using the PCA approach in a wide centrality range of PbPb collisions at 2.76~TeV generated within the HYDJET++ model. The $v^{(1)}_{3}$ results are compared to the $v_{3}\{2\}$ measured by the CMS~\cite{CMS:2013bza} (open green circles) and~\cite{Chatrchyan:2012wg} (open green crosses) and by the ALICE~\cite{Aamodt:2011by} collaborations, and to the $v_{3}\{2,|\Delta\eta| > 2\}$ extracted from the same HYDJET++ simulation using the two-particle correlation method. The error bars correspond to statistical uncertainties.}
\end{center}
\ruledown
\begin{multicols}{2}

\begin{center}
\centering
  \includegraphics[width=0.45\textwidth]{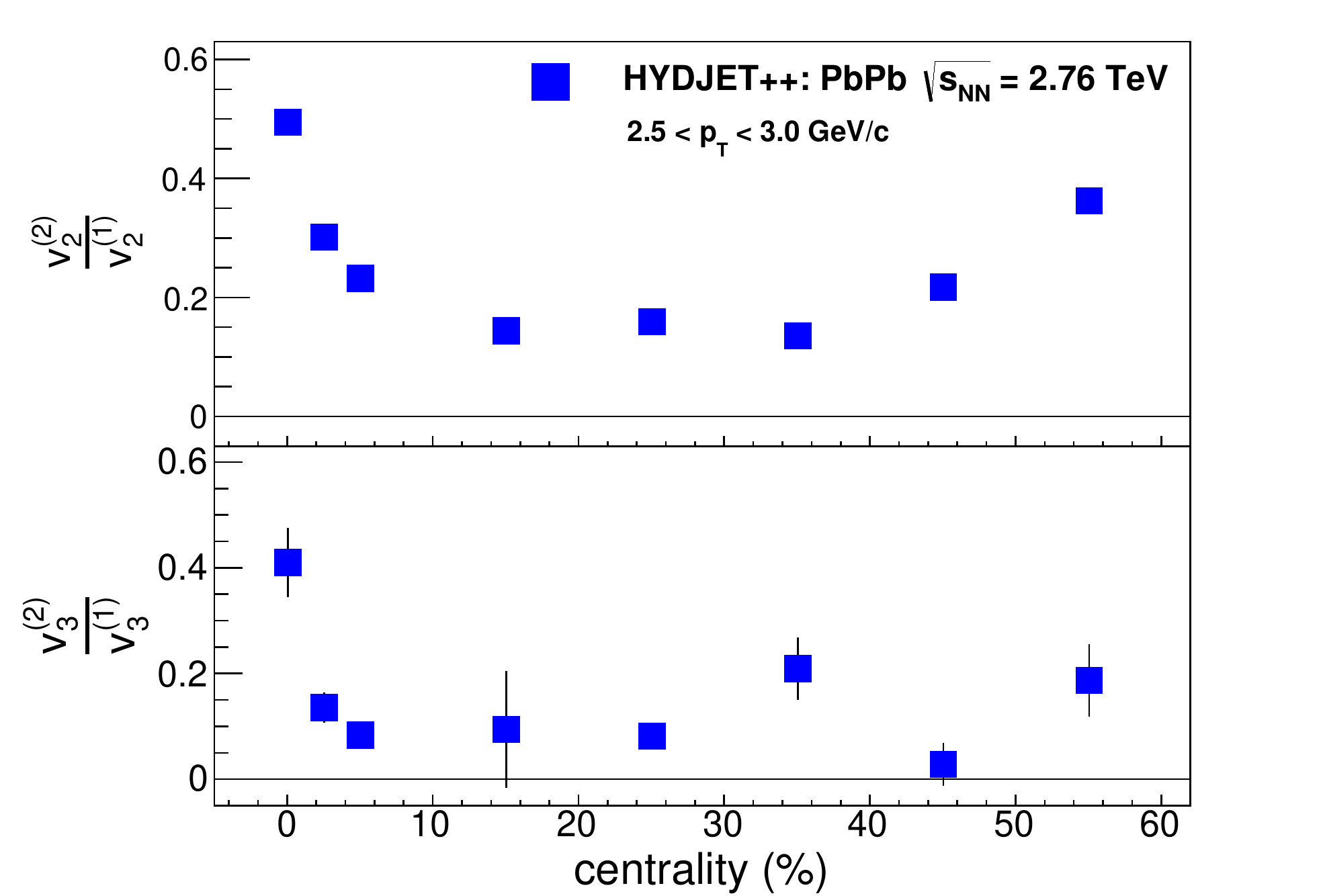}
\figcaption{\label{fig:4} The ratio between values of the sub-leading and leading flow, taken for the highest $p_{T}$ bin, as a function of centrality calculated using the PCA method applied to PbPb collisions at $\sqrt{s_{NN}} =$~2.76~TeV simulated with HYDJET++ event generator. The error bars correspond to statistical uncertainties.}
\end{center}

In order to summarize results, in Fig.~\ref{fig:4} are depicted ratios\footnote{According to Eq.~(2) in~\cite{Mazeliauskas:2015vea}, the connection to the factorization breaking variable is given through ratio $v^{(2)}_{n}/v^{(1)}_{n}$ which gives the relative strength of the effect.} between the sub-leading and leading flow mode. The ratio is calculated from the values taken from 2.5~$< p_{T} <$~3.0~GeV/c where the effect is strongest. The results are presented as a function of centrality. The results in the top panel of Fig.~\ref{fig:4} show that in the case of $n =$~2 the strength of the relative magnitude $v^{(2)}_{2}/v^{(1)}_{2}$ is smallest for events with centralities between 10 and 30\%, i.e. where the elliptic flow is most pronounced. Going to very central collisions, the magnitude of the effect dramatically increases. Also, the effect reaches a significant magnitude going to peripheral collisions. Qualitatively, such behavior is in an agreement with the $r_{2}$ multiplicity dependence presented in~\cite{Khachatryan:2015oea}. Centrality dependence of the ratio which corresponds to the $n =$~3 case is shown in the bottom panel of Fig.~\ref{fig:4}. The $v^{(2)}_{3}/v^{(1)}_{3}$ ratio, integrated over all centralities, is 0.095~$ \pm $~0.009.
As the extracted $v^{(2)}_{3}$ values are small, small fluctuations in their values can easily produce a non-smooth distribution shown in the bottom panel of Fig.~\ref{fig:4}. The overall small $v^{(2)}_{3}$ values found in this analysis are also in a qualitative agreement with the $r_{3}$ multiplicity dependence presented in~\cite{Khachatryan:2015oea}.

\section{Discussion}
\label{sec:dis}
In order to explore the origin of the sub-leading flow signal observed within the HYDJET++ model, beside the analysis of the PbPb data obtained under the 'flow + quenched jets' switch which results are shown in Sect.~4, the pure 'flow' switch has been used for generating PbPb collisions at $\sqrt{s_{NN}} = $~2.76 TeV too. The comparisons between the PCA elliptic and triangular flow results obtained under these two switches are shown in Fig.~\ref{fig:2c} and Fig.~\ref{fig:3c}, respectively. As expected, the pure 'flow' HYDJET++ switch gives a linearly increasing leading flow mode for both $v^{(1)}_{n}$ harmonics $n =$~2 and 3. Also, as expected, the corresponding magnitude, at a given $p_{T}$, is greater with respect to the one extracted from the data obtained under 'flow + quenched jets' switch. The results for the sub-leading flow mode obtained under pure 'flow' switch, contrary to those shown in Sect.~4 are consistent with zero for centralities smaller than 20\%. But, even in the case of the pure 'flow' switch, for centralities above 20\% a modest effect of the sub-leading flow starts to appear. Up to the centrality of 40\% the magnitude of the effect is still smaller with respect to the both, experimental findings from~\cite{HIN-15-010,Milosevic:2016tiw} and from the results obtained using the 'flow + quenched jets' switch. For the most peripheral, 50-60\% the $v^{(2)}_{2}$ magnitude at high enough $p_{T}$ is greater than the experimental one and the one obtained under the 'flow + quenched jets' switch.
\end{multicols}
\ruleup
\begin{center}
  \includegraphics[width=0.8\textwidth]{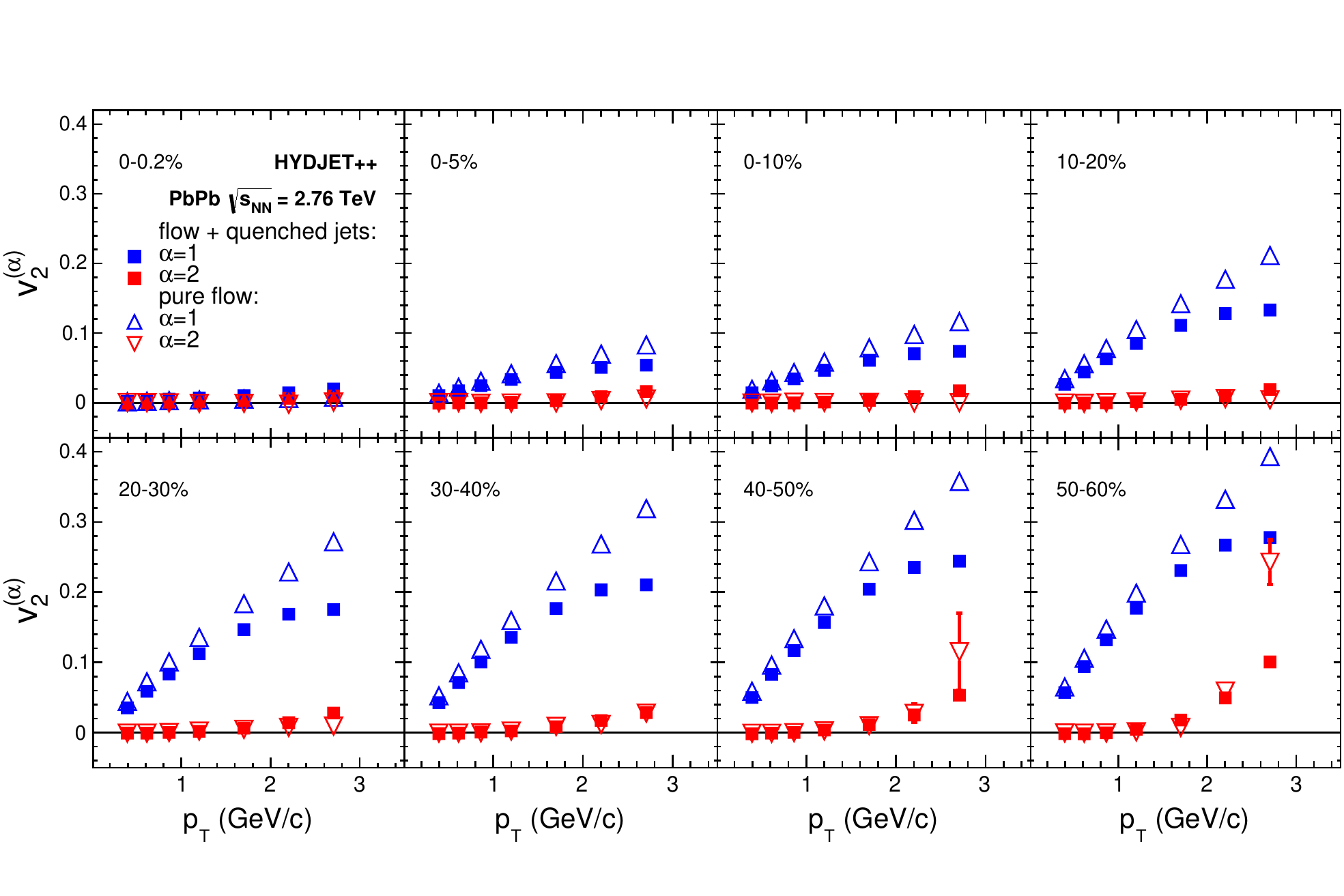}
\figcaption{\label{fig:2c} The leading ($\alpha$ = 1) and the sub-leading ($\alpha$ = 2) flow mode for $n$ = 2 harmonic as a function of $p_{T}$ measured using the PCA approach in a wide centrality range of PbPb collisions at 2.76~TeV generated within the HYDJET++ model under the pure 'flow' switch (triangles) and under the 'flow + quenched jets' switch (squares). The error bars correspond to statistical uncertainties.}
\end{center} 
\ruledown
\begin{multicols}{2}

At first glance, it seems that HYDJET++ data simulated under the pure 'flow' switch should not show existence of the sub-leading flow modes. But, resonance decays and fluctuations of particle momenta together with the topology of peripheral events~\cite{Bravina:2015sda} could imitate hot-spots which at the end could produce a non-zero sub-leading flows. The HYDJET++ data simulated under the 'flow + quenched jets' could have charged pions coming from the jet fragmentation, which due to the interaction with the soft medium and because of different path length with respect to the flow symmetry plane can increase abundance of such high-$p_{T}$ pions near the flow symmetry plane. This also could produce the above mentioned hot-spots and consequently sub-leading flows.

\end{multicols}
\ruleup
\begin{center}
  \includegraphics[width=0.8\textwidth]{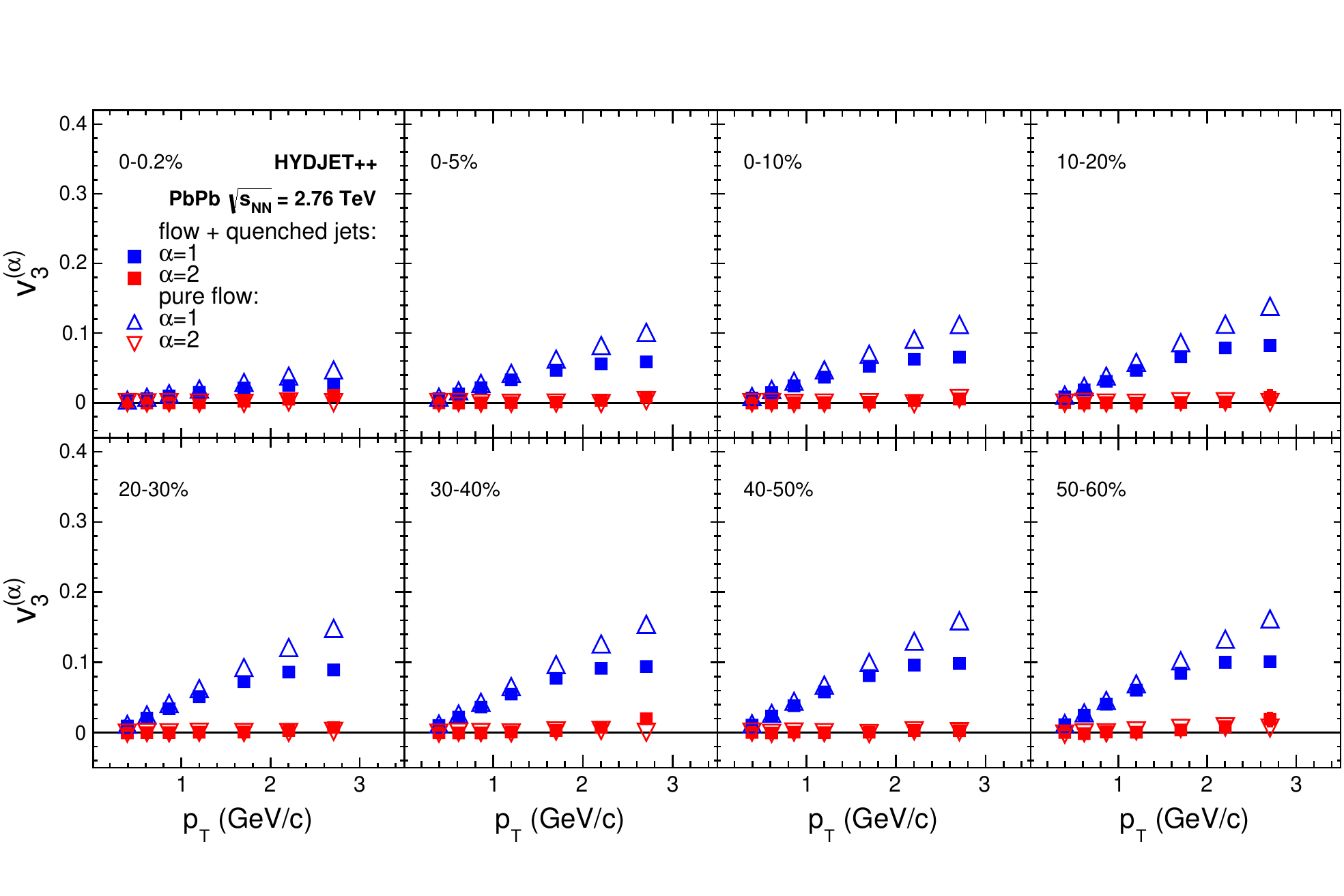}
\figcaption{\label{fig:3c}  The leading ($\alpha$ = 1) and the sub-leading ($\alpha$ = 2) flow mode for $n$ = 3 harmonic as a function of $p_{T}$ measured using the PCA approach in a wide centrality range of PbPb collisions at 2.76~TeV generated within the HYDJET++ model under the pure 'flow' switch (triangles) and under the 'flow + quenched jets' switch (squares). The error bars correspond to statistical uncertainties.}
\end{center}
\ruledown
\begin{multicols}{2}

The results for the sub-leading triangular flow mode are presented in Fig.~\ref{fig:3c}. Similarly to the 'flow + quenched jets' results shown in Sect.~4, the $v^{(2)}_{3}$ values, calculated using the pure 'flow' switch, are close to zero for all centralities and at all $p_{T}$. This again shows that the assumption of the factorization of the two-particle Fourier coefficients into a product of the $v_{3}$ anisotropy harmonics in the case of the pure 'flow' switch is fully valid.

The Pearson coefficient, used to measure the magnitude of factorization breaking is defined~\cite{Bhalerao:2014mua,Khachatryan:2015oea} as
\begin{eqnarray}
\label{pearson}
r_{n}(p^{a}_{T},p^{b}_{T}) = \frac{V_{n\Delta}(p^{a}_{T},p^{b}_{T})}{\sqrt{V_{n\Delta}(p^{a}_{T},p^{a}_{T})V_{n\Delta}(p^{b}_{T},p^{b}_{T})}} \nonumber \\
\sim \langle \cos n(\Psi_{n}(p^{a}_{T}) - \Psi_{n}(p^{b}_{T})) \rangle.
\end{eqnarray}
The $r_{n}$ ratio which is proportional to the cosine term is equal to one if the flow symmetry plane angle is a global quantity. If the factorization breaking occurs then the value of the $r_{n}$ becomes smaller than one. In~\cite{Bhalerao:2014mua} it is shown that the principal component analysis approximates the two-particle Fourier coefficient as
\begin{equation}
\label{pearson-dec}
V_{n\Delta}(p^{a}_{T},p^{b}_{T}) = \sum^{N_{b}}_{\alpha = 1}V^{(\alpha)}_{n}(p^{a}_{T})V^{(\alpha)*}_{n}(p^{b}_{T}),
\end{equation}
where each term in the sum corresponds to a different mode $\alpha$ of the flow fluctuations introduced with Eq.~(\ref{Fmode}). Factorization breaking occurs when non-zero terms with $\alpha \ge $~2 appears in the above sum. Eq.~(\ref{pearson-dec}) is used to reconstruct $V_{n\Delta}$ coefficients from $V^{(\alpha)}_{n}$ flow modes extracted within the principal component analysis. In order to connect the results on the sub-leading flow modes extracted from HYDJET++ generated PbPb collisions at 2.76 TeV with the experimentally seen initial-state fluctuations~\cite{Khachatryan:2015oea}, in Fig.~\ref{fig:7} is shown comparison between the $r_{2}$ and $r_{3}$ ratios, depicted as a function of the transverse momentum difference $p^{a}_{T} - p^{b}_{T}$, measured experimentally in~\cite{Khachatryan:2015oea} and those extracted from HYDJET++ model and calculated using Eq.~(\ref{pearson}) and Eq.~(\ref{pearson-dec}). The comparison is performed only in ultra-central (0-0.2\% centrality) and peripheral (40-50\% centrality) collisions, i.e. where the factorization effect is largest. Using in Eq.~(\ref{pearson-dec}) only the leading and sub-leading flow mode ($N_{b}$ = 2) one observes a fair reconstruction of $r_{n}$ ratios\footnote{The difference in the size of the statistical uncertainties comes because of different statistics used in the experiment and in the HYDJET++ model.}. To improve the reconstruction of $r_{2}$ in ultra-central collisions where the effect of the initial-state fluctuations dominates, one would need to add additional modes ($\alpha \ge $~3) in the two-particle harmonic decomposition. As in the case of the elliptic flow, the sub-leading flow mode corresponding to the triangular flow captures the small factorization effect well.
\ruleup
\begin{center}
  \includegraphics[width=0.45\textwidth]{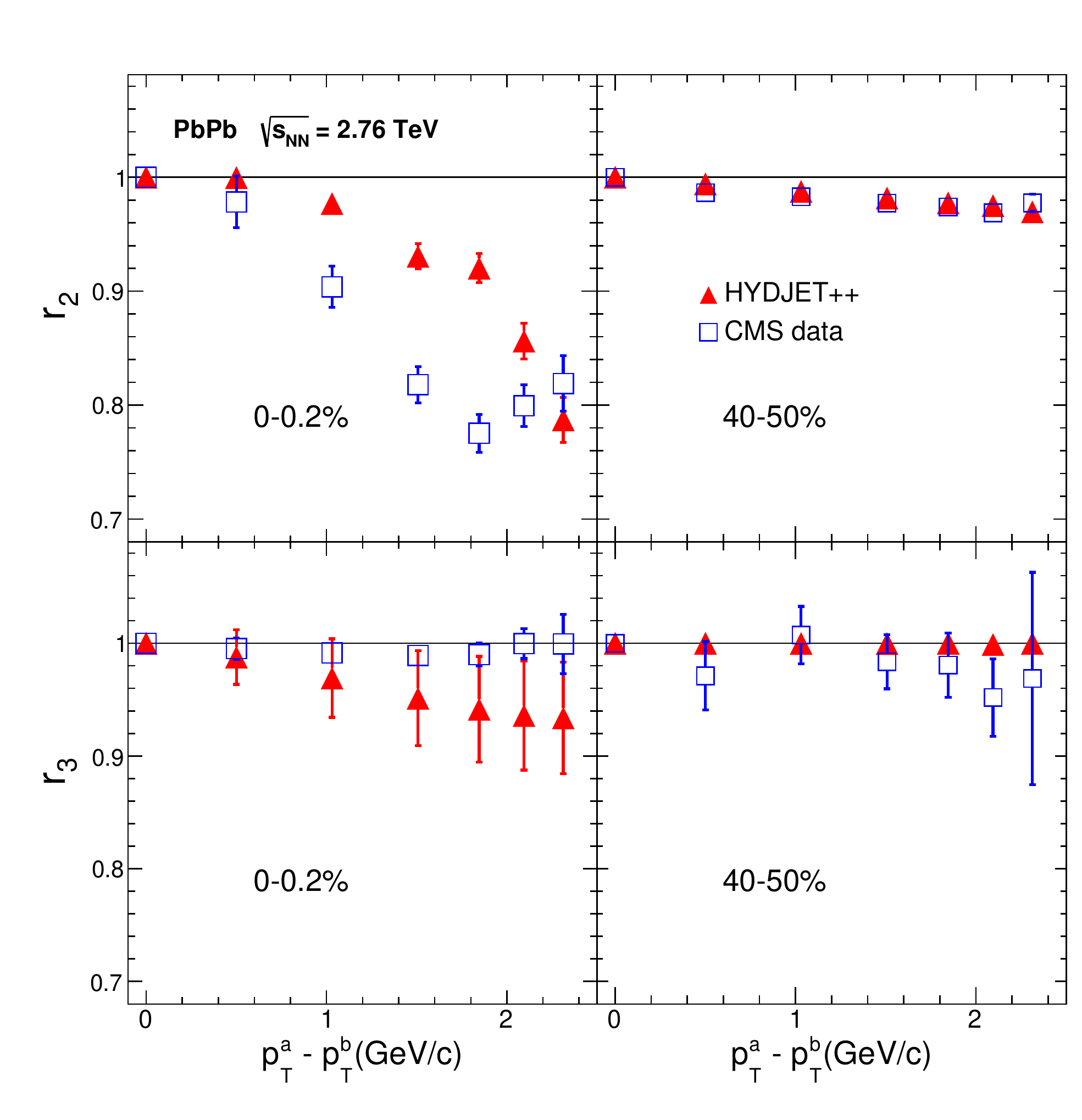}
\figcaption{\label{fig:7}  Comparison of $r_{2}$ (top row) and $r_{3}$ (bottom row) reconstructed with harmonic decomposition using the leading and sub-leading flow mode extracted from HYDJET++ model with the experimental $r_{2}$ and $r_{3}$ values taken from~\cite{Khachatryan:2015oea} for the ultra-central 0-0.2\% and peripheral 40-50\% centrality classes in PbPb collisions at $\sqrt{s_{NN}}$ = 2.76~TeV. The error bars correspond to statistical uncertainties.}
\end{center}
\ruledown

\section{Conclusions}
\label{sec:conc}
The PCA method for studying flow, by its construction, may fully exploits the information contained in the covariance matrix formed from the two-particle Fourier coefficients and thus may provide high sensitivity not only to the standardly defined flow measurements, but also to the influence of the initial-state fluctuations to the hydrodynamic flow. In difference of two-particle correlation method where the information was calculated by integrating over momentum of one of particles which form the pair, within the PCA approach, the detailed information depends on the momenta of both particles of the pair. As the leading flow mode represents the hydrodynamic response to the average geometry, it is essentially equal to the anisotropy harmonics measured using the two-particle correlations method. The sub-leading mode could be understood as the response to the event-by-event initial-state fluctuations which are the main source of the factorization-breaking effect.

The PCA analysis of the PbPb collisions simulated by HYDJET++ model at $\sqrt{s_{NN}} = $~2.76~GeV shows that the leading flow mode, $v^{(1)}_{n}$, for $n = 2, 3$ represents dominant mode and qualitatively describes the experimentally measured $v_{n}$ from two-particle correlations. Additionally, HYDJET++ model also shows existence of the sub-leading flow mode $v^{(2)}_{n}$ which magnitude is in a rather good agreement with the experimental results from the CMS Collaboration. Also, the $r_{2}$ and $r_{3}$ ratios calculated from only leading and sub-leading flow modes extracted from HYDJET++ model data using the principle component analysis fairly reconstructs experimentally measured ratios. This analysis may also provide new insights into the possible influence of the dynamics of the collision onto appearance of the sub-leading flow modes, and help to understand and improve modeling of the evolution of the strongly-coupled quark gluon plasma.

\acknowledgments{The authors would like to thank Igor Lokhtin and his group from Skobeltsyn INP MSU for providing us with HYDJET++ code and useful sugestions.}

\end{multicols}

\vspace{15mm}

\vspace{-1mm}
\centerline{\rule{80mm}{0.1pt}}
\vspace{2mm}

\begin{multicols}{2}

\end{multicols}

\clearpage
\end{CJK*}

\begin{thebibliography}{90}

\vspace{3mm}

\bibitem{Ollitrault:1993ba}
J.-Y. Ollitrault, Phys. Rev. D {\bf 48}: 1132---1139 (1993)

\bibitem{Voloshin:1994mz}
S. Voloshin and Y. Zhang, Z. Phys. C {\bf 70}: 665---672 (1996)

\bibitem{Poskanzer:1998yz}
A. M. Poskanzer and S. A. Voloshin, Phys. Rev. C {\bf 58}: 1671---1678 (1998)

\bibitem{Alver:2010gr}
B. Alver and G. Roland, Phys. Rev. C {\bf 81}: 054905---054913 (2010)

\bibitem{Back:2002gz}
B.B. Back et al. (PHOBOS Collaboration), Phys. Rev. Lett., {\bf 89}: 222301 (2002)

\bibitem{Ackermann:2000tr}
K.H. Adams et al. (STAR Collaboration), Phys. Rev. Lett., {\bf 86}: 402---407 (2001)

\bibitem{Adcox:2002ms}
K. Adcox et al. (PHENIX Collaboration), Phys. Rev. Lett. {\bf 89}: 212301 (2002)

\bibitem{Aamodt:2010pa}
K. Aamodt et al. (ALICE Collaboration), Phys. Rev. Lett. {\bf 105}: 252302 (2010)

\bibitem{ALICE:2011ab}
K. Aamodt et al. (ALICE Collaboration), Phys. Rev. Lett. {\bf 107}: 032301 (2011)

\bibitem{Abelev:2014pua}
B.B. Abelev et al. (ALICE Collaboration), JHEP {\bf 1506}: 190---271 (2015)

\bibitem{Adam:2016izf}
J. Adam et al. (ALICE Collaboration), Phys. Rev. Lett. {\bf 116}: 132302 (2016)

\bibitem{ATLAS:2011ah}
G. Aad et al. (ATLAS Collaboration), Phys. Lett. B {\bf 707}: 330---348 (2012)

\bibitem{ATLAS:2012at}
G. Aad et al. (ATLAS Collaboration), Phys. Rev. C {\bf 86}: 014907---014954 (2012)

\bibitem{Aad:2013xma}
G. Aad et al. (ATLAS Collaboration), JHEP {\bf 11}: 183---240 (2013)


\bibitem{Chatrchyan:2012wg}
S. Chatrchyan et al. (CMS Collaboration), Eur. Phys. J. C {\bf 72}: 2012 (2012)

\bibitem{Chatrchyan:2012ta}
S. Chatrchyan et al. (CMS Collaboration), Phys. Rev. C {\bf 87}: 014902---014936 (2013)

\bibitem{Chatrchyan:2013kba}
S. Chatrchyan et al. (CMS Collaboration), Phys. Rev. C {\bf 89}: 044906---044937 (2014)

\bibitem{CMS:2013bza}
S. Chatrchyan et al. (CMS Collaboration), JHEP {\bf 02}: 088---0126 (2014)

\bibitem{Khachatryan:2015oea}
V. Khachatryan et al. (CMS Collaboration), Phys. Rev. C {\bf 92}: 034911---034937 (2015)

\bibitem{Wang:1991qh}
S. Wang et al., Phys. Rev. C {\bf 44}: 1091---1095 (1991)

\bibitem{Gardim:2012im}
F. G. Gardim, F. Grassi, M. Luzum, and J.-Y. Ollitrault, Phys. Rev. C {\bf 87}: 031901---031906 (2013)

\bibitem{Heinz:2013bua}
U. Heinz, Z. Qiu, and C. Shen, Phys. Rev. C {\bf 87}: 034913---034922 (2013)

\bibitem{Zhou:2014bba}
Y. Zhou, Nucl. Phys. A {\bf 931}: 949---953 (2014)

\bibitem{Bhalerao:2014mua}
R. Bhalerao, J.-I. Ollitrault, S. Pal, and D. Teaney, Phys. Rev. Lett. {\bf 114}: 152301---152306 (2015)

\bibitem{Mazeliauskas:2015vea}
A. Mazeliauskas and D. Teaney, Phys. Rev. C {\bf 91}: 044902---044912 (2015)

\bibitem{Lokhtin:2008xi}
I.P. Lokhtin, L.V. Malinina, S.V. Petrushanko, A.M. Snigirev, I. Arsene, and K. Tywoniuk, Comput. Phys. Commun. {\bf 180}: 779---799 (2009)

\bibitem{Sjostrand:2006za}
T. Sjostrand, S. Mrenna, and P. Skands, JHEP {\bf 0605}: 026---0602 (2006)

\bibitem{Lokhtin:2005px}
I.P. Lokhtin and A.M. Snigirev, Eur. Phys. J. C {\bf 45}: 211---217 (2006)

\bibitem{Mazeliauskas:2015efa}
A. Mazeliauskas and D. Teaney, Phys. Rev. C {\bf 93}: 024913---024928 (2016)

\bibitem{HIN-15-010}
CMS Collaboration, Principal Component Analysis of two-particle azimuthal correlations in PbPb and pPb collisions at CMS, (CERN Document Server, 2015), http://cds.cern.ch/record/2055291. Accessed 27 September 2015

\bibitem{Milosevic:2016tiw}
J. Milosevic for the CMS Collaboration, Nucl. Phys. A {\bf 956}: 308---311 (2016)

\bibitem{Chatrchyan:2013nka}
S. Chatrchyan et al. (CMS Collaboration), Phys. Lett. B {\bf 724}: 213---240 (2013)

\bibitem{Aamodt:2011by}
K. Aamodt et al. (ALICE Collaboration), Phys. Lett. B {\bf 708}: 249---264 (2012
)

\bibitem{Bravina:2015sda}
L. V. Bravina, E. S. Fotina, V. L. Korotkikh, I. P. Lokhtin, L. V. Malinina,E. N. Nazarova, S. V. Petrushanko, A. M. Snigirev, E. E. Zabrodin, Eur. Phys. J. C {\bf 75}, 588---598 (2015)





\end{thebibliography}
\end{document}